\documentclass[prl,twocolumn,showpacs,floatfix,amsmath,amssymb,superscriptaddress]{revtex4-1}
\usepackage{amsfonts}
\usepackage{stmaryrd}
\usepackage{bbm}
\usepackage{mathrsfs}
\usepackage{tipa}
\usepackage{amssymb}
\usepackage{txfonts}
\usepackage{graphicx}
\usepackage{dcolumn}
\usepackage{epstopdf}
\usepackage[colorlinks,linkcolor=blue,urlcolor=blue,citecolor=blue]{hyperref}
\usepackage{multirow}
\usepackage{subfigure}
\usepackage{url}

\begin{document}
\newcommand*{\cm}{cm$^{-1}$\,}
\newcommand*{\Tc}{T$_c$\,}

\title{Optical spectroscopy and ultrafast pump-probe studies on the heavy fermion compound CePt$_2$In$_7$}%force line break\\
\author{R. Y. Chen}
\affiliation{International Center for Quantum Materials, School of Physics, Peking University, Beijing 100871, China}

\author{S. J. Zhang}
\affiliation{International Center for Quantum Materials, School of
Physics, Peking University, Beijing 100871, China}

\author{E. D. Bauer}
\affiliation{Los Alamos National Laboratory, MS E536, Los Alamos, New Mexico 87545, USA}

\author{J. D. Thompson}
\affiliation{Los Alamos National Laboratory, MS E536, Los Alamos, New Mexico 87545, USA}

\author{N. L. Wang}
\affiliation{International Center for Quantum Materials, School of Physics, Peking University, Beijing 100871, China}
\affiliation{Collaborative Innovation Center of Quantum Matter, Beijing, China}

\begin{abstract}
We report optical spectroscopy and ultrafast pump-probe measurements on the antiferromagnetic heavy fermion compound CePt$_2$In$_7$, a member showing stronger two dimensionality than other compounds in CeIn$_3$-derived heavy-fermion family. We identify clear and typical hybridization spectral structures at low temperature from the two different spectroscopy probes. However, the strength and related energy scale of the hybridization are much weaker and smaller than that in the superconducting compounds CeCoIn$_5$ and CeIrIn$_5$. The features are more similar to  observations on the antiferromagnetic compounds CeIn$_3$ and CeRhIn$_5$ in the same family. The results clearly indicate that the Kondo interaction and hybridizations exist in the antiferromagnetic compounds but with weaker strength.
\end{abstract}


\maketitle
\section{introduction}

 Heavy fermion (HF) systems containing elements with a partially filled 4$f$- or 5$f$-electron shell are among the most fascinating materials in condensed matter physics. The strong correlation of $f$-electrons and their coupling to the conduction electrons result in a wide range of exotic phenomena, e.g. magnetic order, unconventional superconductivity, quantum criticality, and a Fermi liquid state with heavy electron mass. Depending on temperature the $f$-electrons show both itinerant and localized behaviors. A great deal of attention has been given to the similarities and peculiarities among different HF systems. Among the most extensively studied systems are the family Ce$_m$M$_n$In$_{3m+2n}$, because different compounds in the family exhibit significantly different ground states in spite of similar lattice structures\cite{Mathur1998,Ikeda2001,Petrovic2007a,Hegger2000,Bauer2010,Macaluso2003}. The compounds are constituted by the alternate stacking of CeIn$_3$ and MIn$_2$ building blocks, where M stands for transition metals. Depending on the specific values of $m$ and $n$, which represent the numbers of CeIn$_3$ and MIn$_2$ layers respectively, there are several types of heavy fermion systems discovered in this family until now: CeIn$_3$\cite{Mathur1998}, CeMIn$_5$\cite{Ikeda2001,Petrovic2007a,Hegger2000}, CePt$_2$In$_7$\cite{Bauer2010}, Ce$_2$MIn$_8$\cite{Macaluso2003} and Ce$_3$MIn$_{11}$\cite{Tursina2013,Kratochvilova2014}, etc.

The ground states of heavy fermion materials generally can be depicted by the Doniach's phase diagram, which was proposed initially to described the competition between Kondo and Ruderman-Kittel-Kasuya-Yosida (RKKY) interactions in cerium compounds\cite{Doniach1977}. Although both of them originate from the exchange interaction $J_K$ between conduction electrons and local $f$ moments, the Kondo effect tends to generate a non-magnetic Fermi liquid ground state with all the local moments screened by conduction electrons and is characterized by the Kondo temperature $T_K\sim exp(-1/J_K)$; whereas, the RKKY exchange interaction enhances long range magnetic order with an energy scale $T_{RKKY} \sim J_K^2$. When $J_K$ is very small, the RKKY interaction dominates the Kondo effect and gives rise to long range magnetic orders. When $J_K$ is sufficiently large, the Kondo effect dominates and the coupling between $f$-electron and conduction electrons will lead to a hybridization gap $\Delta_{HG}$ opening in the density of states (DOS) near the Fermi energy, the energy scale of which is a measure of the hybridization strength. Significantly, a quantum phase transition from magnetic to heavy Fermi liquid state is expected to occur when a delicate energy balance between RKKY and Kondo interactions is approached. Unconventional superconductivity frequently appears in the vicinity of the such quantum phase transition\cite{Monthoux2007}. For example, antiferromagnetic (AFM) order in CeIn$_3$ and CeRhIn$_5$ can be suppressed by the application of pressure and gives way to a heavy fermion superconducting state, because hybridization is strengthened due to a reduction of their lattice constants\cite{Walker1997,Knebel2008}.

As a relative new member of the Ce$_m$M$_n$In$_{3m+2n}$ family,  CePt$_2$In$_7$  also undergoes an AFM phase transition at the Neel temperature $T_N$=5.2 K\cite{growth}. Quantum oscillation measurements reveal the two-dimensionality of its Fermi surfaces, in accord with the layered lattice structure\cite{Altarawneh2011}.
The long range AFM order is reported to be either commensurate or incommensurate by muon-spin rotation \cite{Mansson2014}, Nuclear Quadrupolar Resonance\cite{ApRoberts-Warren2010c} and Nuclear Magnetic Resonance measurements\cite{NMR2011}, depending on details of synthesis conditions for single or polycrystal samples. The AFM transition is suppressed by applying pressure, and superconductivity emerges in a broad dome. Particularly, the highest $T_c$ appears at a critical pressure where the Neel temperature approaches zero Kelvin, implying the existence of a quantum critical point\cite{pressure2013}. In addition, Knight shift measurements reveal that the hybridized quasiparicles begin to "relocalize" before entering the AFM phase, possibly as a precursor to the AFM order\cite{Knight2011}.

Here, we report infrared spectroscopy and ultrafast pump-probe measurements on single crystalline CePt$_2$In$_7$. As these experiments show, infrared spectroscopy is an ideal method in studying heavy fermion materials, since it provides direct information on not only the magnitude of the hybridization gap but also the enhancement of effective mass. The ultrafast measurement also are a very useful tool in detecting small energy gaps in the DOS, which supplies supplementary information on the electronic properties.

\section{infrared spectrocopy}

CePt$_2$In$_7$ single crystals were grown by self-flux method, as described in reference \cite{growth}. Platelike crystals with shiny surfaces were obtained after eliminating extra flux. The in-plane reflectivity $R(\omega)$ was measured by the Fourier transform infrared spectrometer Bruker 113 V, and 80V in the frequency range from 40 to 25 000 \cm. The \textit{in situ} gold and aluminium overcoating techniques were employed to obtain the absolute value of reflectivity. The optical conductivity was derived from $R(\omega)$ through the Kramers-Kronig transformation. The low frequency reflectivity was extrapolated to unit by Hugen-Rubens relations. At the high energy side, $R(\omega)$ was extrapolated as $\omega^{-0.2}$ to 800 000 \cm, and higher frequencies are assumed to decay as $\omega^{-4}$.

\begin{figure}[htbp]
  \centering
  % Requires \usepackage{graphicx}
  \includegraphics[width=7.5cm]{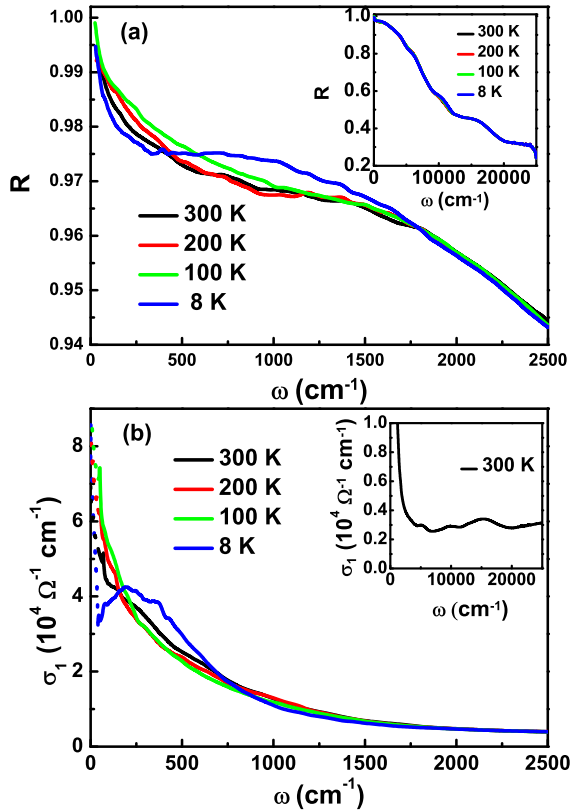}\\
  \caption{(a) The temperature dependent reflectivity $R(\omega)$ of CePt$_2$In$_7$. The inset shows the $R(\omega)$ in a larger frequency range. (b) The temperature dependent optical conductivity $\sigma_1(\omega)$. The inset shows $\sigma_1(\omega)$ in a larger frequency range up to 25 000 \cm.}\label{Fig:ref}
\end{figure}

The frequency dependent reflectivity of CePt$_2$In$_7$ is plotted in Fig.\ref{Fig:ref} (a) at several selected temperatures. Although the resistivity of this compound decreases with decreasing temperature and exhibits a slight downward curvatures around 100 K \cite{growth}, the reflectivity $R(\omega)$ shows a very weak temperature dependence. Except for tiny discrepancies at the dip structure around 5000 \cm and bump around 11 000 \cm, the reflectivity at different temperatures almost overlap, as shown in the inset of Fig.\ref{Fig:ref} (a). This result suggestes that the electronic structures of the compound are quite stable against temperature changing. The main panel of Fig.\ref{Fig:ref} (a) displays the enlarged view of $R(\omega)$ in the frequency range below 2500 \cm, which focuses on the low energy evolutions. When the frequency decreases to zero, $R(\omega)$ approaches unity at all measured temperatures, indicative of metallic behavior. Above 100 K, the low energy $R(\omega)$ increases slightly with cooling, consistent with a typical metallic temperature dependence. However, $R(\omega)$ exhibits a very weak depletion structure at 8 K, demonstrating the renormalization of underlying electronic structures.
Remarkably, this feature is quite similar with the reflectivity of antiferromagnetic CeRhIn$_5$ \cite{optical2005}; whereas, there are significant contrasts with that of superconductors CeCoIn$_5$ and CeIrIn$_5$ \cite{optical2005,Singley2002}.

The real part of the optical conductivity $\sigma_1(\omega)$ of CePt$_2$In$_7$ is displayed in Fig.\ref{Fig:ref} (b). For temperatures above 100 K, the low energy responses are dominated by a Drude-like shape. The half width of the Drude peak narrows slightly as temperature decreases, and the dc conductivity ($\sigma_1(\omega=0)$) is enhanced concurrently, in good agreement with previous transport measurement \cite{growth}. When temperature is lowered to 8 K, a minimum appears at very low frequency, and a weak bump develops that corresponds to a suppression of the reflectivity. At the same time, the Drude peak shifts dramatically to lower energy. All these characters are consistent with the typical optical response of heavy fermion materials. Upon entering the coherence state where the local moments start to become screened by conduction electrons, the scattering rate is sharply reduced and gives rise to a narrowing of the Drude peak. Concurrently, spectral weight of the Drude component is substantially removed because the effective mass of free carriers becomes much heavier. The weak bump feature is attributed to optical excitations across a hybridization gap.

\begin{figure}[htbp]
  \centering
  % Requires \usepackage{graphicx}
  \includegraphics[width=7.5cm]{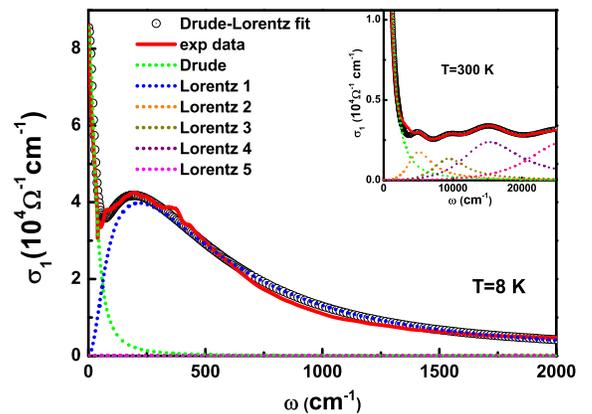}\\
  \caption{The experimental optical conductivity $\sigma_1(\omega)$ along with the decomposed Drude and Lorentz components of CePt$_2$In$_7$ at 8 K in the main panel and 300 K in the inset.}\label{Fig:conductivity}
\end{figure}

In order to analyze the optical conductivity more quantitatively, we use the Drude-Lorentz model to decompose $\sigma_{1}(\omega)$:
\begin{equation}\label{Eq:DL}
\epsilon(\omega)= \epsilon_{\infty}-\frac{\omega_{p}^2}{\omega^2+i\omega/\tau_{D}}+ \sum_{j}{\frac{S_j^2 }{\omega_j^2-\omega^2-i\omega/\tau_j}}. %\label{chik}
\end{equation}
Here, $\varepsilon_{\infty}$ is the dielectric constant at high energy; the middle term is the Drude component that characterizes the electrodynamics of itinerate carriers, and the last term is the Lorentz component that describes excitations across energy gaps or interband transitions. We find that $\sigma_1(\omega)$ can be well reproduced by one Drude term and four Lorentz terms at high temperatures, but an additional Lorentz term is required at 8 K, as shown in Fig.\ref{Fig:conductivity}. Some temperature dependent fitting parameters are listed in Table \ref{1}.

\begin{table}[bhtp]
\setlength\abovecaptionskip{0.5pt}
\caption{Fitting parameters of $\sigma_1(\omega)$ for different temperatures. At 8 K, a new Lorentz term centered at 215 \cm is added. \label{1}}
\vspace{-1em}
\begin{center}
\renewcommand\arraystretch{1.5}
\begin{tabular}{p{1.5cm} p{1.5cm} p{1.2cm} p{1.2cm} p{1.2cm} p{1.1cm}}
\hline
\hline
 T&$\omega_p$&$\gamma_D$&$\omega_1$&$\gamma_1$&$S_1$\\
\hline
300 K&39580&613&&&\\
200 K&39580&575&&&\\
100 K&39580&488&&&\\
8 K&12080&30&215&593&38000
\\
\hline
\hline
\end{tabular}\\
\end{center}
\end{table}

Among these fitting parameters, the square of the plasma frequency $\omega_{P}^2$ is proportional to $n/m^{*}$, where $n$ is the density of free carriers and $m^{*}$ stands for the effective mass of quasiparticles. $\omega_{P}$ is essentially a constant for temperatures above 100 K, but it drops sharply at 8 K. Assuming the carriers density $n$ does not change with temperature, the effective mass can be estimated to be $m^{*}=10.7 m_b$, where $m_b$ is the corresponding band mass. This value is a little bit higher than estimated from quantum oscillation measurement\cite{Altarawneh2011}.
Furthermore, $\gamma_D=1/\tau_{D}$ is the scattering rate of free carriers, which is an order of magnitude larger at 100 K than 8 K. The large $\gamma_D$ at high temperature is attributed to strong scattering of conduction electrons by Ce 4$f$ local moments that ultimately are screened by free carriers in the coherent state and act like itinerate quasiparticles.

For the Lorentz terms, $\omega_j$, $\gamma_i=1/\tau_j$ and $S_j$ represent the resonance frequency, width and square root of the oscillator strength, respectively. There are four nearly temperature independent Lorentz term appearing at high energies, which stem from interband excitations. We only display parameters for the additional Lorentz term emerging at the lowest temperature, which is associated with the formation of hybridization gap. The central frequency $\omega_1=215$ \cm ($\sim$ 27 meV) can be considered as the energy scale of the hybridization gap. Compared with other members in the Ce$_m$M$_n$In$_{3m+2n}$ family, it is much smaller than that of CeCoIn$_5$ and CeIrIn$_5$ but is a little bit larger than for CeRhIn$_5$ and CeIn$_3$ \cite{optical2005,Singley2002,Lee2008,Iizuka2012}. Moreover, the hybridization peak feature in the optical conductivity of CePt$_2$In$_7$  is much weaker than in CeCoIn$_5$ and CeIrIn$_5$ but is comparable with that in CeRhIn$_5$ and CeIn$_3$. According to Doniach's model, the non-magnetic ground states of CeCoIn$_5$ and CeIrIn$_5$ indeed suggest stronger hybridization strength, and the relative weaker character of CePt$_2$In$_7$ is in accord with its antiferromagnetic ordered ground state, which reflects relatively weaker Kondo coupling in comparison with the RKKY interactions between local moments being mediated by conduction electrons.

\section{ultrafast pump probe measurement}

Because the hybridization gap structure detected by infrared spectroscopy is quite weak, we also performed time resolved pump-probe measurements on the same sample. This measurement is very sensitive to the presence of low energy gaps and provides useful supplementary information. A Ti:sapphire oscillator was utilized as the source of both pump and probe beams and produces 800-nm laser pulses with 100 fs duration at a 80 MHz repetition rate. The pump beam was modulated with a frequency of 1 MHz and polarized perpendicular to the probe beam. The pump fluence is about 2 $\mu$J/cm$^2$, 10 times stronger than the probe pulses.

\begin{figure}[htbp]
  \centering
  % Requires \usepackage{graphicx}
  \includegraphics[width=7.5cm]{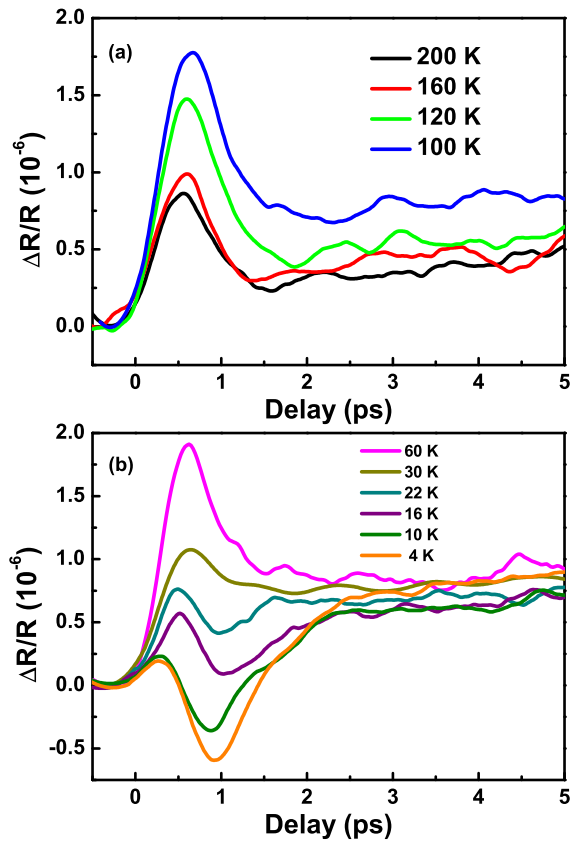}\\
  \caption{The photoinduced transient reflectivity $\Delta R/R$ at several selected temperatures. The data are smoothed for clarity.}\label{Fig:delta}
\end{figure}

Figure \ref{Fig:delta} shows the photoinduced reflectivity change $\Delta R/R$ as a function of time delay at high (a) and low (b) temperatures.
The transient reflectivity is initially enhanced due to thermomodulation of the pump pulse that excites electrons to empty states much higher than the Fermi energy and leaves holes below. Then $\Delta R/R$ relaxes back to a flat background within several picoseconds. Above 24 K, this relaxation can be well described by a single exponential decay $\Delta R/R$ = $A$ exp$(-t/\tau_1)$, where $A$ stands for the amplitude of the photoinduced reflectivity and $\tau_1$ is the relaxation time of the decay. The flat background cannot recover to equilibrium within our measurement range of up to several hundreds of picosecond, so it is most likely corresponding to a heat diffusion process that usually takes several nanoseconds. It is clearly seen in Fig.\ref{Fig:delta} (a) that at high temperatures the magnitude of $\Delta R/R$ increases monotonically with temperature cooling, whereas the relaxation time is almost temperature independent. This is a typical response of conventional metals and can be interpreted by the phenomenological two-temperature model. In this model, the electrons are first heated by the pump pulses and establish a quasi-equilibrium state in tens of femtoseconds. Thus, the transient temperature of the electron system $T_e$ becomes much higher than the lattice temperature $T_L$. These hot electrons transfer their excess energy to the lattice via slower electron-phonon interactions, and the relaxation time of this process is directly dependent on the electron-phonon coupling constant $\lambda$.

Between the temperature range 100 K $>$ $T$ $>$ 60 K, we find that $\Delta R/ R$ remains nearly unchanged. With temperature further deceasing below 60 K, both $A$ and $\tau_1$ drop rapidly upon cooling. Moreover, an additional decay channel with opposite sign is clearly observed below 24 K, which provides strong evidence for the renormalization of the electronic structure. Here, two exponential functions are required to fit the relaxation: $\Delta R/R$ = $A$ exp($-t/\tau_1$) + $B$ exp($-t/\tau_2$).
We have extracted the relaxation time of the second decay channel from an exponential fitting. This relaxation time increases monotonically with decreasing temperature and grows to be almost 6 times longer at the lowest temperature. The slowing down of the relaxation time also has been demonstrated in many other heavy fermion compounds, such as CeCoIn$_5$\cite{Demsar2006}, Yb$_{1-x}$Lu$_x$Al$_3$\cite{Demsar2009}, URu$_2$Si$_2$ \cite{Liu2011b} and so on\cite{Nair2012,Qi2013}. In these materials, enhancement of the decay time is attributed to the opening of hybridization gaps, which can be well explained by the Rothwarf-Taylor (RT) model\cite{Rothwarf1967}.
Below the Kondo temperature, the photoexcited quasiparticles relax initially through electron-electron and electron-phonon scattering to states just above the hybridization gap, usually within subpicosecond time scale. The further combination of electron and hole pairs across $\Delta_{HG}$ generates high frequency phonons (HFP) with energy $\omega_{HFP}\geq \Delta_{HG}$, which subsequently break more electron-hole pairs. This avalanche process severely impedes the recovery of quasiparticles back to the equilibrium states, leading to the well known phonon bottleneck effect that substantially increases the relaxation time.

\begin{figure}[htbp]
  \centering
  % Requires \usepackage{graphicx}
  \includegraphics[width=7.5cm]{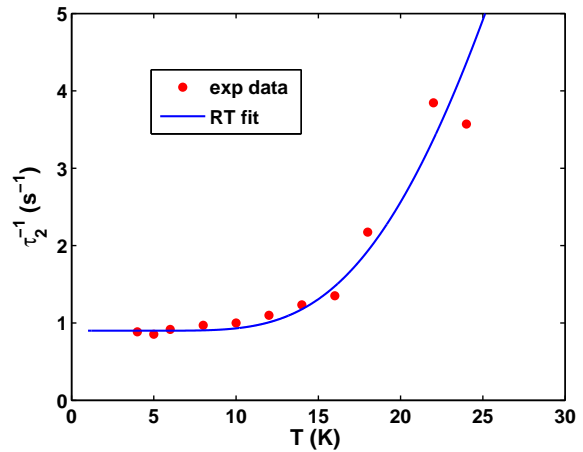}\\
  \caption{The temperature-dependent relaxation time of the second decay channel. The blue solid curve is the result of fitting to the RT model.}\label{Fig:tau2}
\end{figure}

According to the RT model, the temperature dependence of the relaxation time follows\cite{Kabanov2005,Demsar2006}:
\begin{equation}\label{Eq:tau}
  \tau^{-1}(T)=\Gamma[\delta(\epsilon n_T+1)^{-1}+2n_T],
\end{equation}
where $\Gamma$, $\delta$ and $\epsilon$ are all temperature independent fitting parameters and $n_T$ represents the density of thermally excited quasiparticles which is related to the energy scale of the low energy gap $\Delta_{g}$ via
$n_T\simeq \sqrt{T} exp(-\Delta_{g}/2T)$. We find that $\tau_2$ follows Eq. \ref{Eq:tau} quantitatively and the gap energy is determined to be $\Delta_{g}$=140 K ($\sim$ 13 meV), as shown by the solid curve in Fig. \ref{Fig:tau2}. Note that $\Delta_{g}$ is much smaller than $\Delta_{HG}$ identified by infrared measurements. Because optical spectroscopy is responsive to direct charge excitations, whereas pump probe techniques are sensitive to indirect interband transitions, the latter is very likely to yield a smaller gap energy. For example, similar results also are revealed in the heavy fermion superconductor PuCoGa$_5$ \cite{Talbayev2010}.

Another remarkable fact is that the monotonic increase of the amplitude $A$ upon cooling terminates around 100 K, which is far away from the direct observation of two exponential decays. It is worth noting that the temperature dependent resistivity also shows a downwards bending around 100 K. Therefore, we propose that the hybridization gap begins to develop just below this particular temperature,
and the dropping of $A$ and $\tau_1$ is actually ascribed to the emergence of the additional decay channel whose amplitude ($B$) is opposite that of $A$. Upon lowering temperature, both $B$ and $\tau_2$ should increase due to the opening of the hybridization gap, although it is too weak to be observed in the presence of relative large $A$ and $\tau_1$ above 24 K.
In contrast, the relaxation dynamics of CeCoIn$_5$ follow a single exponential decay, and the initial electron-phonon coupling process is buried in the phonon bottleneck signals\cite{Demsar2006}. Moreover, the relaxation time across the hybridization gap in CeCoIn$_5$ is enhanced by two orders of magnitude upon cooling, much larger than in CePt$_2$In$_7$. All these characters suggest that the hybridization strength of CePt$_2$In$_7$ is rather weak, consistent with the infrared measurement.

\section{summary}

In conclusion, we have performed infrared and ultrafast pump probe measurements on the antiferromagnetic heavy fermion compound CePt$_2$In$_7$. Upon entering the heavy fermion coherence state, the optical conductivity reveals a very weak hybridization gap feature with an energy of 27 meV. Concurrently, the Drude component shifts dramatically to lower frequency because of the screening of local moments and the spectral weight of it is substantially removed due to enhancement of the effective mass. Ultrafast pump-probe reveals an additional relaxation channel accompanied with the development of the hybridization gap. The relaxation time of this channel increases upon cooling until the lowest temperature. This response agrees well with the expected behavior of heavy electrons relaxing across a hybridization gap. The energy scale of this indirect gap is extracted to be 13 meV, much smaller that the direct gap identified by infrared measurement. Although the presence of a hybridization is  demonstrated clearly in CePt$_2$In$_7$, its strength is much weaker than in superconducting CeCoIn$_5$ and CeIrIn$_5$ but comparable with that in the antiferromagnets CeIn$_3$ and CeRhIn$_5$. The results imply that the AFM ground state of CePt$_2$In$_7$ can be ascribed to its relatively weak hybridization strength.

\begin{center}
\small{\textbf{ACKNOWLEDGMENTS}}
\end{center}

This work was supported by the National
Science Foundation of China (11120101003, 11327806), and the 973
project of the Ministry of Science and Technology of China
(2012CB821403). Work at Los Alamos was performed under the auspices of the U.S. Department of Energy, Office of Basic Energy Sciences, Division of Materials Sciences and Engineering.

\bibliographystyle{apsrev4-1}
  \bibliography{CePt2In7}

\end{document}